# Engineering Decisions in MBSE: Insights for a Decision Capture Framework Development

*Nidhal Selmi, Jean-Michel Bruel, Sébastien Mosser, Matthieu Crespo, Alain Kerbrat*

**Abstract**

Decision-making is a core engineering design activity that conveys the engineer's knowledge and translates it into courses of action. Capturing this form of knowledge can reap potential benefits for the engineering teams and enhance development efficiency. Despite its clear value, traditional decision capture often requires a significant amount of effort and still falls short of capturing the necessary context for reuse. Model-Based Systems Engineering (MBSE) can be a promising solution to address these challenges by embedding decisions directly within system models, which can reduce the capture workload while maintaining explicit links to requirements, behaviors, and architectural elements. This article discusses a lightweight framework for integrating decision capture into MBSE workflows by representing decision alternatives as system model slices. Using a simplified industry example from aircraft architecture, we discuss the main challenges associated with decision capture and propose preliminary solutions to address these challenges.

## 1. Introduction

In response to the increasing complexity in engineered systems, Model-Based Systems Engineering has emerged as a solution to better manage this complexity by enhancing traceability, consistency, and collaboration. In this context, MBSE promises to help engineering teams work more efficiently by shortening development times through the use of overarching single-source-of-truth models (Campo et al.). However, while traditional MBSE focuses on "what" the system is and how it functions through structural and behavioral modeling, there is little focus on "why" a certain solution is chosen. This article proposes insights on a value-based decision capture framework that can further shorten rework and revisiting time. Decisions translate the engineer's knowledge —whether explicit, implicit, or tacit—into courses of action. This makes them a valuable asset for engineering teams, as they help consolidate and leverage engineering knowledge for future projects. The remainder of this article is structured as follows. Section 2 illustrates the limitations of unstructured decision communication through a simplified example. Section 3 presents a preliminary proposal for the decision capture framework and discusses how it can potentially address the identified challenges. Finally, we conclude with a concise summary of our findings and outline promising directions for future research.

## 2. Problem Statement and Challenges

### 2.1. Motivating Example: An Illustration of Unstructured Decision Capture

To highlight the need for a proper, systematic approach to managing engineering decisions in the industry, we will use a simplified industry example: the implementation of a cabin depressurization function in an overall aircraft design problem. Once the airplane has landed, especially after an emergency landing, the cabin can have some residual differential pressure compared to the ambient pressure. This can lead to significant hazards: opening the aircraft doors in the presence of a pressure differential can be impossible or damage the door structure.
Engineers with different roles in the design team are involved in treating this problem: the aircraft architect, who supports trade-off studies at the aircraft level, suggests two possible alternatives. The first option is to allocate this function to the doors system: the opening and pressure relief mechanisms are connected directly, resulting in a reliable mechanical link and reduced system complexity. The second option is to extend the existing control system to implement this function. Although this increases control complexity, it reduces weight and cost per door. The performance specialist evaluates the impact of both choices on the overall performance of the aircraft. Once evaluated, the overall aircraft architect opts for option one and communicates this choice to the system designer, who refines the architecture based on this choice.

Although the decision-making process and roles of the involved engineers are well-defined, the capture of the decision information has been done in an unstructured manner: the overall weight and performance evaluation carried out by the performance specialist is stored on their machine. The pros and cons of each alternative were communicated in a team meeting and documented in the meeting minutes. The resulting allocation of the function is stored as part of the MBSE model. Later on, if the allocation of the depressurization function is questioned by another member of the design team, they would need to collect this decentralized information from the engineers involved. Especially when developing highly complex, long lifecycle systems like aircraft, this is not straightforward. The responsible engineers might be elsewhere by the time the information is needed. Additionally, the large number of documents, analysis results, and meeting minutes stored in an unstructured manner makes it nearly impossible to find the right information.

Therefore, it is clear that there is potential value in using a more systematic approach to capturing decisions, along with their rationale and analysis. The potential value of such an approach is closely tied to the promised development cost-reduction benefits of model-based systems engineering. Making explicit the traceability between the "whats" and their respective "whys" can make the revisiting, challenging, and understanding of past decisions much more seamless. Another potential benefit is the consolidation of engineering knowledge inside the organization. For instance, reusing decisions based on context similarity can minimize rework time, leading to more efficient development.

To summarize, the decision management approach we aim to implement should focus on reducing the difficulties of revisiting and understanding the reasons behind design results and provide more structured decision information that can be useful for analysis and reuse.

### 2.2. Main Challenges in Decision Capture

A series of informal discussions was conducted with Airbus engineers to elicit their needs and operational constraints. Based on these discussions and reviews of existing literature (Regli et al., Lee, Harrison et. al.), we identified the following challenges:
- The effort to capture decisions along with their relevant information can be obtrusive to the engineer's main activity: Separately documenting the reasons for design artifacts results in additional workload.
- Access to decision information has to be straightforward and intuitive.
- The context of the decision-making activity is very important for understanding and reusing past decisions in future design problems.
- To ensure an efficient, value-based capture of decisions, it is necessary to identify which decisions are key: A compromise needs to be found to capture only decisions that will likely be questioned, revisited, or reused.

Further challenges include scalability, confidentiality, granularity, and complexity of decisions. These have not been considered in this first proposal and will be investigated in future iterations. Another goal of our research is to leverage past decisions, including their contextual information, to formalize and enhance design knowledge, which is a valuable asset for organizations and design teams. Additionally, we aim to continuously verify the properties of decisions and related artifacts to ensure they follow defined guidelines, e.g., to ensure the completeness of the decision information.

## 3. Strategic Directions for a Decision Capture Framework

In this section, we will discuss a preliminary proposal for a decision management framework utilizing MBSE model slices. A high-level overview of the proposed decision management framework is illustrated in Fig. 1. The main approach is to capture the results of decisions as slices of the system's descriptive model. The primary assumption underlying this approach is that all engineering decisions have a direct impact on the MBSE model. Making explicit this link between the MBSE model artifacts and the underlying decision information has two main benefits: (i) capture and access efficiency and (ii) potential consolidation of knowledge through context elicitation.

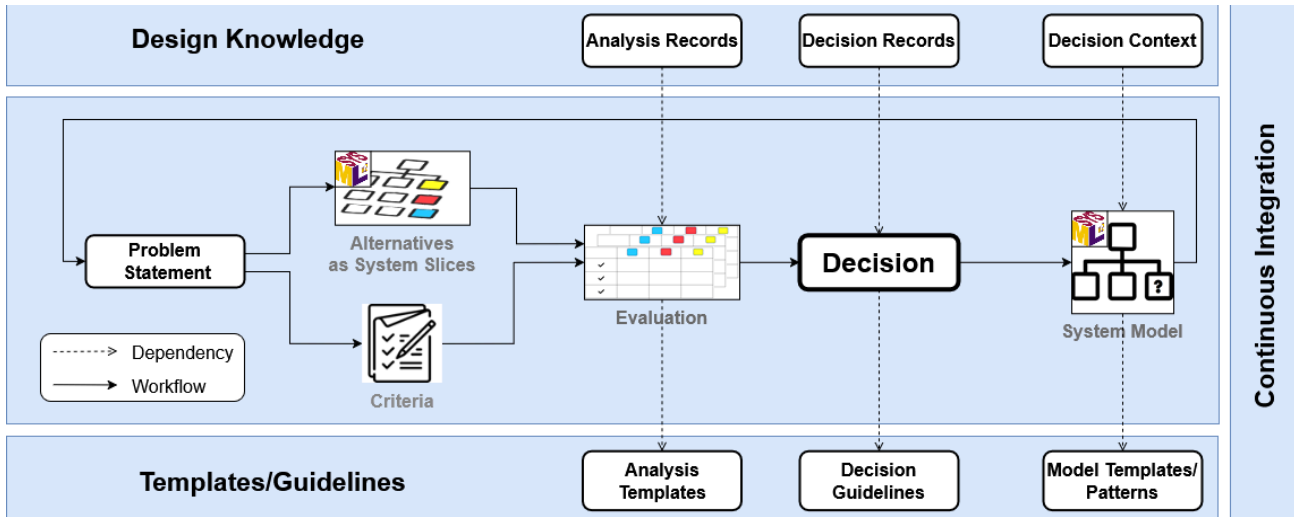

*Figure 1. Decision Management Framework Overview*

We will use the example from Section 2 to illustrate how these benefits can be achieved. Both proposed solutions to implement the depressurization function can be identified as subsets of the MBSE model. In this case, the resulting artifact of either alternative is simply an allocation link between the depressurization function and either the cabin door component or the pneumatic and air systems component. The system modeler captures both alternatives in the MBSE model along with the evaluation criteria. Here, it is essential to note that decision rationales can also be strategic rather than merely a technical evaluation. In this example, the performance specialist creates a traceability link to their evaluation results, i.e., analysis, test, and simulation cases, assuming that these results are stored as part of the digital thread and that links can be created between them and the MBSE artifacts. The chosen course of action is simply recorded by labeling one of the alternatives as the preferred option. The capture is more efficient and less obtrusive as the system modeler captures the decisions in the same environment as the descriptive model. If the new team member needs to access the rationale for choosing the allocation to the door system instead of the control system, they can simply navigate to the rationale and disregard alternatives through the allocation link between the activity diagram and the door system. This way of accessing the rationale seems most intuitive, as we first observe what was created, to then ask why and why not.

Different types of interdependencies can be identified to better contextualize decisions. Once defined, these relationships can allow us to analyze the impact of changing one decision on the remaining linked decisions. The ISO 42010 standard for Software, Systems, and Enterprise Architecture provides examples of the possible relationships between decisions (ISO). These relationships include constraints, influences, enables, triggers, forces, and subsumes, among others. For instance, the decision to use door vent flaps could trigger a new design problem concerning the positioning of these vents. Here, the "triggers" relationship is defined as a chronological link between decisions to create a streamlined decision-making process, directing the architect to the next recommended focus point (Zimmermann et al.). Creating these dependencies between decisions can help contextualize a single decision, which is essential for understanding and reuse. Additionally, since the alternatives are captured as slices of the MBSE model, their context within the model and their content regarding artifacts can offer valuable information for categorizing and contextualizing the decision.

4. **Conclusion and perspectives**

Decisions are the translation of engineers' knowledge into courses of action. This knowledge represents an essential asset for engineering teams and organizations. Not only does value-based capture of decisions render the design process more efficient and reduce revisiting times, but it can also potentially help consolidate organizational memory and better leverage design knowledge for future projects. Our approach to capturing decision outcomes as slices of the MBSE model to overcome intrusiveness issues. Another potential benefit is the contextualization of decisions within the system and the design process, which can help gain a better understanding of the rationale and potentially improve knowledge consolidation.

**Acknowledgments**

The work presented in this paper has been supported by the CoCoVaD Industrial Chair, which is funded by Airbus to the Toulouse Jean Jaurès University.

**About the authors**

**Nidhal Selmi** is a first-year PhD student at Toulouse Jean Jaurès University and IRIT. His research aims to improve and integrate decision management in MBSE frameworks.

**Jean-Michel Bruel** is a professor at the University of Toulouse and a member of the IRIT CNRS laboratory. He has been the holder of the Chair of Model-Driven Systems Engineering between AIRBUS and Toulouse Jean Jaurès University since 2022.

**Sébastien Mosser** is an Associate Professor of Software Engineering in the Department of Computing and Software at McMaster University and a member of the McMaster Centre for Software Certification.

**Matthieu Crespo**, PhD in computer science and fluid dynamics, is an MBSE expert at Airbus Commercial Aircraft and the modeling & simulation architect for the ZEROe zero-emission program.

**Alain Kerbrat**, PhD in computer science, is an aircraft architect on Airbus's DDMS transformation project. He defines MBSE specifications, combining models and textual requirements to streamline supplier exchanges and support early verification and validation of system architectures.